\documentclass[a4paper, amsfonts, amssymb, amsmath, reprint, showkeys, nofootinbib, twoside,superscriptaddress,prb]{revtex4-1}
\usepackage[english]{babel}
\usepackage[utf8]{inputenc}
\usepackage[colorinlistoftodos, color=green!40, prependcaption]{todonotes}
\usepackage{amsthm}
\usepackage{mathtools}
\usepackage{physics}
\usepackage{xcolor}
\usepackage{graphicx}
\usepackage[left=23mm,right=13mm,top=35mm,columnsep=15pt]{geometry} 
\usepackage{adjustbox}
\usepackage{placeins}
\usepackage[T1]{fontenc}
\usepackage{lipsum}
\usepackage{csquotes}
\usepackage[version=4]{mhchem}
\usepackage{siunitx}  

\usepackage[pdftex, pdftitle={Article}, pdfauthor={Author}]{hyperref} % For hyperlinks in the PDF
\DeclareSIUnit\formulaunit{f.u.}
\DeclareSIUnit\atom{atom}
\newcommand{\comment}[1]{}
\usepackage{xcolor}

\usepackage[capitalize]{cleveref}
\usepackage{svg}
\crefname{subsection}{subsection}{subsections}
\begin{document}
\title{Accurate effective harmonic potential treatment of the high-temperature cubic phase of Hafnia}

\author{Sebastian Bichelmaier}
\affiliation{Institute of Materials Chemistry, TU Wien, A-1060 Vienna, Austria}
\affiliation{KAI GmbH, Europastrasse 8, A-9524 Villach, Austria}
\author{Jesús Carrete}
\affiliation{Institute of Materials Chemistry, TU Wien, A-1060 Vienna, Austria}
\author{Michael Nelhiebel}
\affiliation{KAI GmbH, Europastrasse 8, A-9524 Villach, Austria}
\author{Georg K. H. Madsen}
\email[Correspondence email address: ]{georg.madsen@tuwien.ac.at}%
\affiliation{Institute of Materials Chemistry, TU Wien, A-1060 Vienna, Austria}

\date{\today} 

\begin{abstract}
\ce{HfO_2} is an important high-$\kappa$ dielectric and ferroelectric, exhibiting a complex potential energy landscape with several phases close in energy. It is, however, a strongly anharmonic solid, and thus describing its temperature-dependent behavior is methodologically challenging.
We propose an approach based on self-consistent, effective harmonic potentials and higher-order corrections to study the potential energy surface of anharmonic materials. The introduction of a reweighting procedure enables the usage of unregularized regression methods and efficiently harnesses the information contained in every data point obtained from density functional theory. This renders the approach highly efficient and a promising candidate for large-scale studies of materials and phase transitions. 
We detail the approach and test it on the example of the high-temperature cubic phase of \ce{HfO_2}. Our results for the thermal expansion coefficient, $\alpha_V \approx \SI{3.3e-5}{\per\kelvin}$, are in agreement with existing experimental ($\approx \SI{4\pm 1 e-5}{\per\kelvin}$) and theoretical ($\approx \SI{5\pm1e-5}{\per\kelvin}$) work. Likewise, the bulk modulus agrees well with experiment. We show the detailed temperature dependence of these quantities. 

\end{abstract}


\maketitle

\section{Introduction} \label{sec:introduction}
One of the biggest drawbacks of density functional theory (DFT) calculations is the lack of temperature-induced effects. However, due to the exponential growth in computing power and continuous methodological developments, the previously prohibitively expensive calculations necessary to remedy that shortcoming are becoming viable for the investigation of new materials. Consequently, the inclusion of temperature is a prominent theme in many current computational efforts \cite{xie_PRB_1999,grabowski_PRB_2007,souvatzis_PRL_2008,grabowski_PRB_2009,hellman_PRB_2011,hellman_PRB_2013,roekeghem_CPC_2021,ehsan_PRB_2021}.
    
In the present study we focus on hafnia, \ce{HfO_2}, which has a multi-faceted phase diagram \cite{tobase_PRL_2018} %,batra_jpc_2017} 
and numerous industrially relevant applications, ranging from a high-$\kappa$ gate dielectric for semiconductors in its amorphous \cite{wilk_JAP_2001} and (more recently suggested)  tetragonal phase \cite{fischer_JAP_2008}, to a ferroelectric in its orthorhombic states \cite{boescke_APL_2011} for e.\,g.\ nonvolatile memory applications \cite{mueller_JSS_2015}. 

An accurate DFT treatment of its temperature-dependent behavior has proven difficult to achieve in previous theoretical efforts \cite{huan_PRB_2014}. In the simplest approach, the effect of temperature is included by means of the harmonic approximation (HA), where the phonon modes of the system are described as independent harmonic oscillators. The second-order interatomic force constants (IFCs) are obtained by applying small displacements and mapping the corresponding forces induced by them. However, in the case of structures governed by anharmonic potential energy surfaces (PES), the HA might yield imaginary frequencies, thus indicating mechanical instability, even when experiments confirm the existence of those structures.
    
Ab-initio molecular dynamics (AIMD) approaches \cite{Iftimie_PNAS2005} can, in principle, treat such temperature-stabilized structures, but obtaining the free energy of reasonably complex systems through thermodynamic integration proves to be a resource-intensive task and quickly becomes intractable. Moreover, AIMD treats the nuclear motion in a completely classical fashion, and therefore cannot capture effects such as zero-point motion, which can be relevant for e.g. accurately describing the vibrations of light and strongly bonded atoms.

An emerging category of alternatives to AIMD can be labelled as effective harmonic potentials (EHP). The idea goes back to 1955 \cite{hooton_PM_1955} and in essence involves determining the best HA to the part of the PES which dominates nuclear motion. Temperature dependent contributions to the free energy are then included using independent harmonic oscillators based on these EHPs. EHPs have proven to be a rich starting point for understanding temperature dependent behavior using ab-initio methods and, as computational power has become available, prompted various implementations and formulations of the underlying theory \cite{souvatzis_PRL_2008,hellman_PRB_2013,errea_PRL_2013, Tadano_PRB15,Stern_PRB16,roekeghem_CPC_2021,Monacelli_JPCM21}. The implementations mainly differ in how the PES is sampled with methods including stochastic sampling and molecular dynamics trajectories as well as how the deviation between the EHP and the PES is accounted for.
    
In the present study we focus on the high-temperature cubic (Fm$\bar{\textnormal{3}}$m) hafnia phase (c-\ce{HfO_2}). c-\ce{HfO_2} is an example of a temperature stabilized structure where the small displacement HA yields imaginary frequencies \cite{huan_PRB_2014}. We show how the use of reweighting in combination with unregularized regression can be employed to obtain the temperature-dependent EHP. Furthermore, the correction for the deviation between the EHP and the true DFT PES is discussed in detail. We compare our results to the available AIMD calculations and experiments.

\section{Method} \label{sec:method}
\subsection{Background}\label{subsection:bg}
A system described by a Hamiltonian $\hat{\mathcal{H}}$ is in a state of thermal equilibrium at constant volume, temperature and number of particles when its free energy 

\begin{equation}
    \mathcal{F}[\hat{\rho}] = \mathrm{tr}\left(\hat{\rho} \hat{\mathcal{H}} \right) + T \mathrm{tr}\left(\hat{\rho}\log\hat{\rho}\right)%=E-TS,  
    \label{eqn:free_ene}
\end{equation}
is at a minimum. This equilibrium state is described by a particular quantum mechanical density matrix, $\hat{\rho}$, which, were it known, would provide access to the whole thermodynamics of the system. However, it is impossible for all but trivial model systems to solve this problem exactly. 

The EHP can be formulated as a variational problem \cite{Errea_PRB14,Monacelli_JPCM21} where a trial density matrix, $\hat{\tilde{\rho}}$, which exactly solves a corresponding trial Hamiltonian, $\hat{H}$, is introduced. $\hat{H}$ differs from the true Hamiltonian $\hat{\mathcal{H}}$ only by the form of the potential energy operator $\hat{V}$, as opposed to $\hat{\mathcal{V}}$. Minimizing the free energy with respect to the trial density matrix is guaranteed by the Gibbs-Bogoliubov inequality \cite{Isihara_JPGP68} to provide an upper bound on the free energy
\begin{equation}
     \mathcal{F}[\hat{\rho}] \leq F[\hat{\tilde{\rho}}] = \mathcal{F}[\hat{\tilde{\rho}}] + \mathrm{tr}\left[\hat{\tilde{\rho}}(\hat{\mathcal{V}} - \hat{V})\right]=\mathcal{F}[\hat{\tilde{\rho}}]+{F_{\mathrm{corr}}}, 
     \label{eqn:gibbs}
\end{equation}
In the harmonic approximation, the trial potential is parametrized as

\begin{equation}
    \mel{\vec{u}}{\hat{V}}{\vec{u}} = V(\vec{u}) =  \sum_{ij}\frac{1}{2} u_i \Phi_{ij} u_j, 
    \label{eq:v}
\end{equation}
in terms of the displacements from the minimum-energy configuration, $\vec{u}$, and the second-order force constants, $\Phi_{ij}$, whose eigenvalues and eigenvectors
\begin{equation}
    \Phi_{ij} = \sum_\lambda \omega_\lambda^2 \epsilon_{\lambda i} \epsilon_{\lambda j}^*,
    \label{eqn:fc}
\end{equation}
will hereafter be denoted by $\omega_\lambda^2$ and $\vec{\epsilon}_\lambda$. The indices $i,j$ denote the ions and the Cartesian directions. 
Within the harmonic approximation, the projection onto real space of the trial density matrix can be expressed in closed form

\begin{equation}
    \mel{\vec{u}}{\hat{\tilde{\rho}}}{\vec{u}} = \tilde{\rho}(\vec{u}) = \frac{1}{\sqrt{(2\pi)^{3N}}\sqrt{\left|C\right|}} e^{-\frac{1}{2}\vec{u}C^{-1}\vec{u}}.
    \label{eqn:dm}
\end{equation}
The covariance matrix $C$ can be obtained from the aforementioned $\omega_\lambda$ and $\vec{\epsilon}_{\lambda}$ through a well-known result of quantum statistical mechanics

\begin{equation}
    C_{ij}= \frac{\hbar}{2\sqrt{M_iM_j}}\sum_\lambda \frac{1}{\omega_\lambda \tanh{\frac{\hbar \omega_\lambda}{2k_BT}}}\epsilon_{\lambda i} \epsilon_{\lambda j}^\dagger,
    \label{eqn:covmatrix}
\end{equation}
where $M$ correspond to the masses of the ions.  Furthermore, the expression for the free energy, $\mathcal{F}[\hat{\tilde{\rho}}]$, is given by

 \begin{equation}
     \mathcal{F}(T) = \sum_\lambda \hbar \omega_\lambda \left( \frac{1}{2} + k_\textnormal{B}T\log{\left[1-\exp{-\frac{\hbar \omega_\lambda}{k_\textnormal{B} T}}\right]}\right).
     \label{eqn:vib}
 \end{equation}
$\mathcal{F}$ depends directly on the temperature $T$ and indirectly on the harmonic trial potential through $\omega_\lambda$, \cref{eqn:fc}. The optimal trial potential thus depends on the temperature. Ignoring $F_\text{corr}$, \cref{eqn:gibbs}, and the  temperature-dependence of the effective potential  results in the well-known quasi-harmonic approximation.

\subsection{Temperature-dependent effective potentials}\label{subsection:free_ene}

We implement the search for the optimal EHP by approximating the real-space density matrix by means of canonical importance sampling and treating the interdependence of $\mathcal{F}$ and $\Phi$ as a self-consistent problem. When self consistency is reached, this corresponds to minimizing $F[\hat{\tilde{\rho}}]$ \cite{roekeghem_CPC_2021}.

The starting point is the second-order force-constant matrix, and corresponding potential  $V^{(1)}$, obtained through small displacements as implemented in Phonopy \cite{togo_SM_2015}. From the eigenvalues and eigenvectors  and the temperature of interest the associated trial density, $\tilde{\rho}^{(1)}$, is obtained through \cref{eqn:dm,eqn:covmatrix}. We replace the imaginary square roots of possible negative eigenvalues from intermediate steps with their modulus. From this probability density the first set of displacements, $\mathcal{S}^{(1)}$, are drawn. A new EHP is obtained by calculating the potential energies and forces corresponding to the displacements using DFT and finding the parametrization of the force constants in \cref{eqn:fc} which best represent the relationship between forces and displacements, as will be discussed below. The iterative process then progresses by contructing a new density matrix using \cref{eqn:covmatrix,eqn:dm}. To aid convergence, the new trial density matrix, $\tilde{\rho}^{(k)}$, is obtained through a Pulay mixing scheme \cite{pulay_jcc_1982} with a memory of $n=5$ and a mixing parameter of $\alpha=0.1$, as commonly used. A new set of displacements is now drawn and the process continues until convergence is reached. 

To efficiently use all the data obtained from DFT, the reweighting factor \cite{errea_PRL_2013} is introduced
\begin{equation}
        w^{(g\rightarrow k)}_m = \frac{\tilde{\rho}^{(k)}(\vec{u}^{(g)}_m)}{\tilde{\rho}^{(g)}(\vec{u}^{(g)}_m)},
    \label{eq:weight}
\end{equation}
Thereby displacement vectors, $\vec{u}^{(g)}_m$, belonging to a set drawn in a previous iteration, $\mathcal{S}^{(g)}$, and their corresponding forces and potential energies can be included as if they belong to the current set, $\mathcal{S}^{(k)}$. 

\begin{figure}
    \centering
    \includegraphics{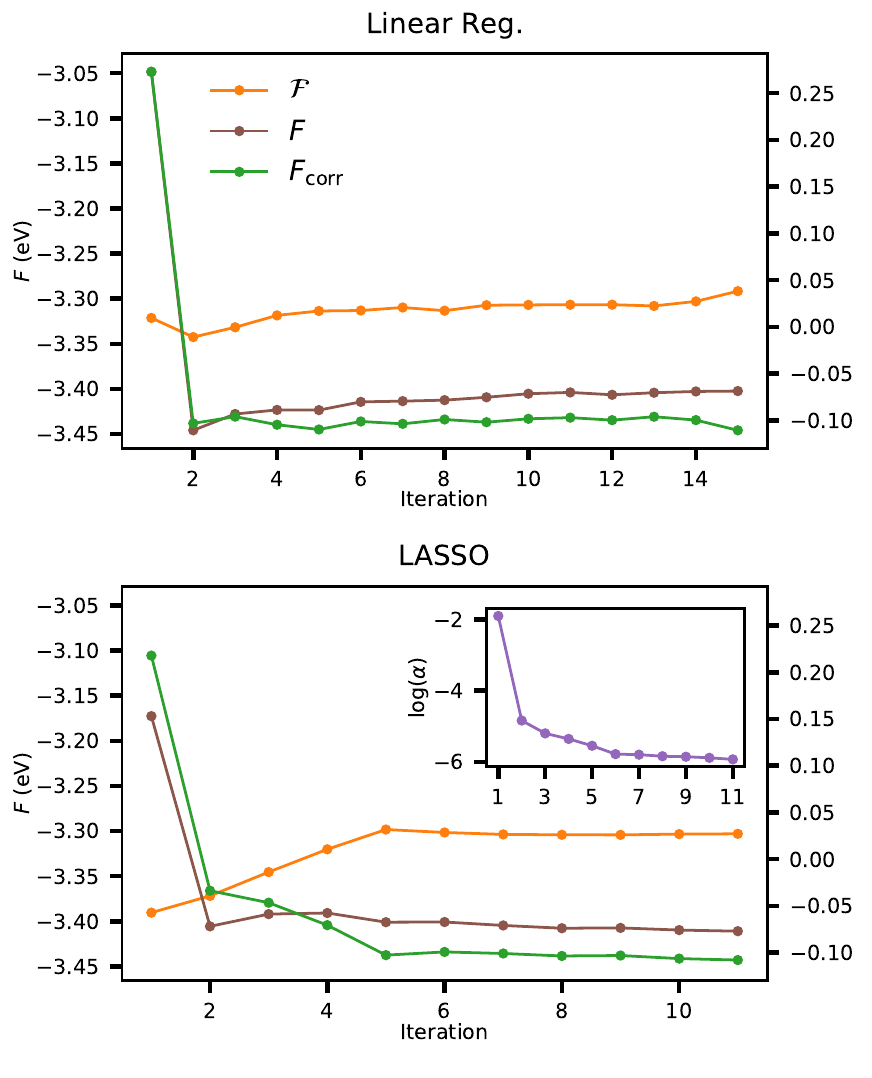}
    \caption{Convergence of the free energy contributions with iterations for the exemplary case of the \SI{0}{\kelvin} equilibrium structure, $V_0 = \SI{32.62}{\angstrom^3}$, at $T=\SI{2500}{\kelvin}$. The upper plot is for a least squares penalty function and the lower for  LASSO. The total free energy, $F$, and the harmonic contribution, $\mathcal{F}$, are given w.r.t. the y-axis to the left and $F_\textrm{corr}$ with respect to the y-axis to the right.  In every iteration 5 new structures are added.}
    \label{fig:convergence_Lasso_Linreg}
\end{figure}

Using the reweighting factors significantly increases the amount of available data and allows using an unregularized fitting procedure to obtain the force constant matrix.  The trial potential for iteration $k$ is found  by finding the force constants which minimize the weighted sum of the least-squares deviations from the calculated forces,
i.\,e.\,,
\begin{equation}
    \sum_g\sum_m w^{(g\rightarrow k)}_m \left\lVert \vec{f}^{(g)}_{m} + \Phi \vec{u}^{(g)}_{m} \right\rVert_2^2
    \label{eqn:lstsq}
\end{equation}
In \cref{fig:convergence_Lasso_Linreg} the free energy evaluated according to \cref{eqn:vib} is shown as a function of the iterations until convergence for the \SI{0}{\kelvin} equilibrium volume using a temperature of $T=\SI{2500}{\kelvin}$. Typically the convergence criterion of $\Delta\mathcal{F}<\SI{2.5}{\milli\electronvolt\per\formulaunit}$ is reached in $10 - 15$ iterations when 5 structures are added per iteration for the initial temperature point, totaling 50 - 75 DFT runs.  

Alternatively, a penalty 
\begin{equation}
    \bigg( \sum_g\sum_m w^{(g\rightarrow k)}_m \left\lVert \vec{f}^{(g)}_{m} + \Phi \vec{u}^{(g)}_{m} \right\rVert_2^2+\alpha \left\lVert \Phi \right\rVert_1\bigg)
    \label{eqn:lasso}
\end{equation}can be defined. This is known as the LASSO minimization target function.  The $\alpha$ parameter determines the strength of the $L_1$ regularization and promotes sparsity of the force constant matrix, which can be a computational advantage \cite{Fransson_NPJCM20}.
By partitioning the data into five complementary subsets and minimizing \cref{eqn:lasso} for a range of $\alpha$-values, the regularization strength can be tuned to provide the model which generalizes best, i.e. has the best performance on the subsets not used for obtaining the force constants. This so-called 5-fold cross validation procedure \cite{scikit-learn} is performed at every step of the iteration. 
In \cref{fig:convergence_Lasso_Linreg} we compare the convergence behaviour at $T=\SI{2500}{\kelvin}$ for the \SI{0}{\kelvin} equilibrium structure. Notably, the $\alpha$ parameter quickly approaches zero as the calculation progresses. 
This is understandable as regularization is typically applied when fitting samples which insufficiently cover the sample space. Thus, as the amount of data points available for fitting increases, every fold in the cross validation procedure will be more and more equally representative of the rest, so the $L_1$ penalty will actually hinder minimizing \cref{eqn:lasso} and will be forced towards zero by the algorithm itself, effectively resulting in \cref{eqn:lstsq}.

We have chosen the unregularized least-squares approach, \cref{eqn:lstsq} for two reasons: We are, for all but the first few iterations, confronted with an overdetermined system, as only a total of $52$ independent force constants remain after considering symmetry and the cut-off and every sample provides $576$ force-displacement pairs. Furthermore, if a non-zero $L_1$-penalty was indeed used throughout the calculation, the force constants obtained would not fulfill the property of minimizing the free energy once self-consistency is reached \cite{roekeghem_CPC_2021}. As argued above, the regularization parameter must approach zero, because the coverage increases with every iteration. While we would presumably not have arrived at an artificially increased free energy, using LASSO only provided a minor speed up, while introducing additional uncertainty in the results. 

The correction term, $F_{\mathrm{corr}}$, has previously been calculated by representing the DFT PES by a simpler form \cite{errea_PRL_2013,Monacelli_JPCM21} or by using the trajectory obtained from AIMD \cite{hellman_PRB_2013,Metsanurk_PRB19}. We  calculate $F_{\mathrm{corr}}$  directly from the DFT potential energies obtained from the same sampling as used for determining the trial EHP, \cref{eqn:lstsq}, as a weighted average
\begin{equation}
    F_{\mathrm{corr}} = \frac{1}{W}\sum_{g}\sum_{m} w^{(g \rightarrow k)}_m \big(\mathcal{V}(\vec{u}^{(g)}_m) - V^{(k)}(\vec{u}^{(g)}_m)\big) ,
    \label{eqn:correction}
\end{equation}
where $W$ is the sum of all the weights. Similar to the EHP the reweighting allows all DFT calculations to be used for obtaining $F_{\mathrm{corr}}$ and \cref{fig:convergence_Lasso_Linreg} illustrates that the convergence is also comparable, meaning that convergence of the total free energy, $F$, is reached within $10-15$ iterations.

\begin{figure}
    \centering
    \includegraphics{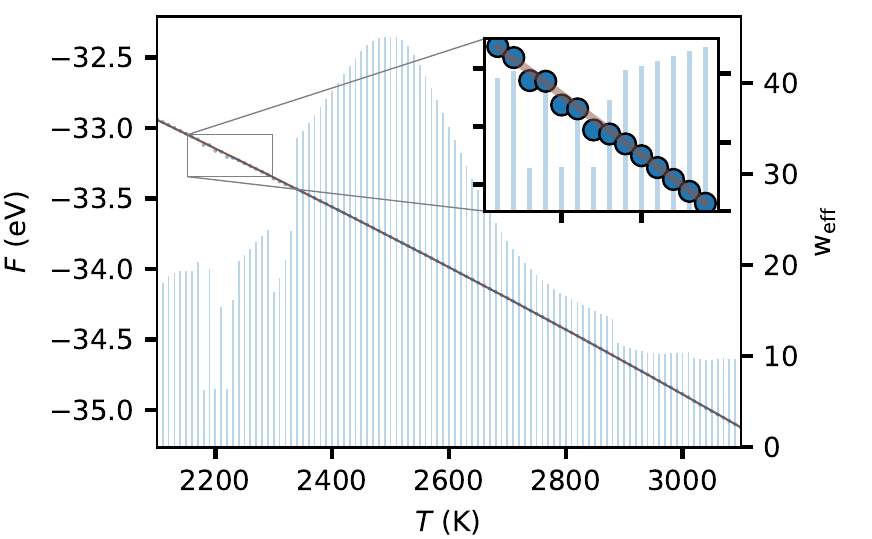}
    \caption{Free energy and effective samples as a function of temperature at $+\SI{2}{\percent}$ deformation, i.e. a volume of $\SI{34.62}{\angstrom^3}$. In regions where the effective samples, as shown by the bars in the background, are low, the free energy shows discontinuous behaviour. The brown line is a smoothing spline using the effective samples as weights, whereas the blue points are the data points.}
    \label{fig:effective_samples_spline}
\end{figure}

 It is straightforward to extend the formalism described above to reuse samples drawn at temperature $T_1$ for a different temperature $T_2$, by building reweighting factors accounting for this. This provides a fast way to calculate force constants at temperatures near $T_1$. At temperatures more different from $T_1$ the sample set might not be adequate anymore, necessitating augmentation by additional DFT runs. As an example we mention that using the displacements and forces obtained for $T_1 = \SI{2500}{\kelvin}$, \cref{fig:convergence_Lasso_Linreg}, at $T_2 = \SI{2100}{\kelvin}$, but reweighted according to \cref{eq:weight} results in convergence after adding only two additional iterations. Once convergence has been obtained for a mesh of temperatures, free energies can be obtained at intermediate temperatures without additional DFT calculations. As a measure for when additional calculations are necessary, we use the effective number of samples,
\begin{equation}
    w_\text{eff}^{(k)} = \frac{\Big(\sum_g\sum_m w^{(g\rightarrow k)}_m\Big)^2}{\sum_g\sum_m \big(w^{(g \rightarrow k)}_m\big)^2}.
    \label{eqn:weff}
\end{equation}
As can be observed in in \cref{fig:effective_samples_spline}, the number of effective samples is indeed an excellent metric for the trustworthiness of the data at a given temperature. However, it would be grossly inefficient to augment the data if the poorly-sampled regions are small and constrained and the surrounding points are described by the existing samples well-enough. To prevent this, we apply a smoothing spline weighted with the effective samples to $F(T)$. This procedure avoids artifactual oscillations in later results. 

\subsection{Computational details}
The DFT calculations were performed using the Vienna Ab Initio Simulation Package (VASP)  \cite{kresse_PRB_1996,bloechl_PRB_1994,kresse_PRB_1999}, where we utilized the Perdew-Berke-Ernzerhof (PBE) exchange and correlation potential \cite{perdew_PRL_1996} along with an energy cutoff of \SI{600}{\electronvolt}. The force calculations for the phonons were performed in a $4\times4\times4$ supercell using just the $\Gamma$-point. We used Phonopy \cite{togo_SM_2015} with the non-analytical correction described in \cite{Wang_2010} for obtaining the initial small displacements force constants.  The descriptors for fitting the force constants [\cref{eqn:lstsq} and \cref{eqn:lasso}] are obtained using scikit-learn \cite{scikit-learn} and the cluster formalism established in Ref.~\onlinecite{eriksson_advtheosimul_2019} given a pre-defined cutoff $r_{cut}=\SI{7}{\angstrom}$.

We performed the steps outlined above for various deformations of the \SI{0}{\kelvin} equilibrium structure; we included volumes from $V=$\SIrange{30.70}{37.77}{\angstrom^3\per\formulaunit}, as well as various temperatures ranging from \SIrange{2100}{3100}{\kelvin}. To arrive at a simple but general analytical expression we then fit the free energies using,

\begin{equation}
    F(V;T) = c_0 + \frac{c_1}{V^{1/3}} + \frac{c_2}{V^{2/3}} + \frac{c_3}{V},
    \label{eqn:SJEOS}
\end{equation}
achieving an average deviation between actual and fitted free energy of less than \SI{7}{\milli\electronvolt\per\formulaunit}, which corresponds to less than $\SI{3}{\milli\electronvolt\per\atom}$. This allows for finding the equilibrium lattice parameter and volume at every temperature point of interest with a high accuracy.

\section{Results} \label{sec:sectionIII}
We settled on a temperature range from \SIrange{2100}{3100}{\kelvin} to ensure full coverage of the stable region of c-\ce{HfO_2}, \SIrange{2800}{3100}{\kelvin} \cite{wang_ACS_2006}, with the upper bound close to the melting temperature. To accurately treat this temperature range, we performed an initial self-consistent run at $T_1 =\SI{2500}{\kelvin}$, augmented it with samples at $T_2 = \SI{2100}{\kelvin}$ and, guided by the effective sample size, \cref{eqn:weff}, included $T_3 = \SI{3000}{\kelvin}$ for some deformations.

\begin{figure}[h]
    \centering
    \includegraphics{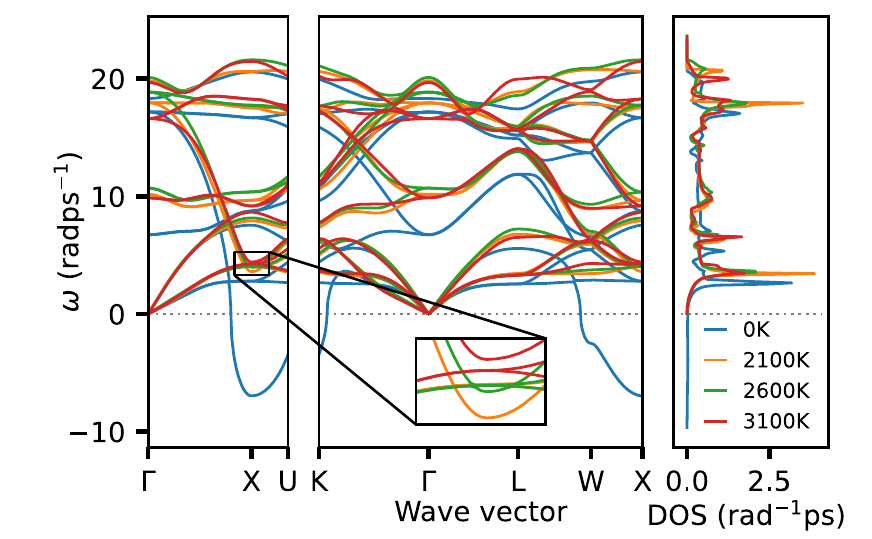}
    \caption{Phonon band structure and density of states of c-\ce{HfO_2} at the \SI{0}{\kelvin} equilibrium volume. The soft mode at \SI{0}{\kelvin} indicated by the negative frequencies at \textit{X} in blue continuously hardens as temperature increases and the structure becomes stable.}
    \label{fig:band_structure}
\end{figure}

As an example, we show the phonon band structure of the \SI{0}{\kelvin} equilibrium structure in \cref{fig:band_structure}, as obtained using small displacements and using the EHP at elevated temperatures. 
As can be seen c-\ce{HfO_2} shows an instability in the small displacements (\SI{0}{\kelvin}) phonon spectrum at $X=(0,1/2,1/2)$ in the Brillouin zone, which, in the structurally very similar \ce{ZrO_2} has been linked to the cubic-to-tetragonal phase transition \cite{kuwabara_PRB_2005}. 
 Within the studied temperature range, we see a continuous hardening (shown in the inset) of said mode indicating that temperature-induced anharmonic effects are stabilizing this phase. 
For the additional volumes that were studied (\SIrange{-6}{16}{\percent} volume changes) we find a similar behaviour and can report stable phonon spectra for the whole volume and temperature range. 

\begin{figure}[h]
    \centering
    \includegraphics{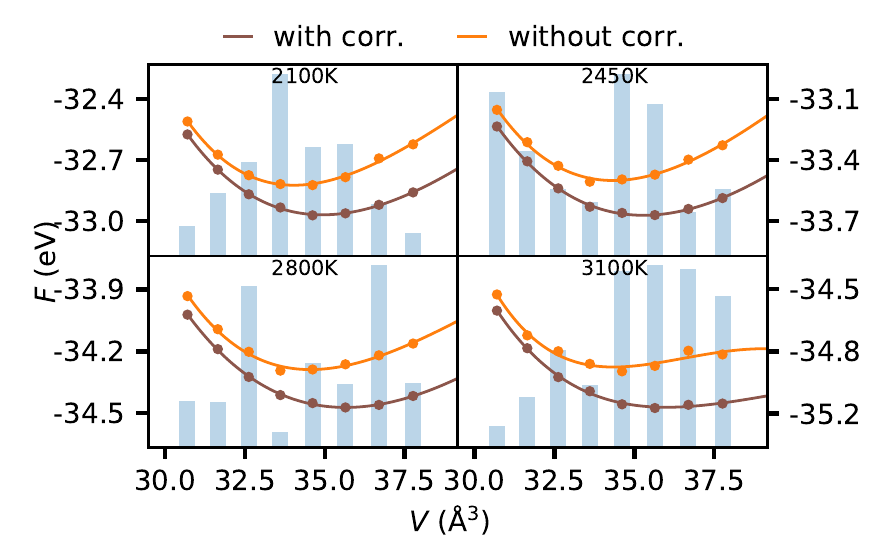}
    \caption{Comparison of the free energies obtained for various volumes and temperatures with and without the correction. The solid lines are the fitted SJEOS as described in \cref{eqn:SJEOS}, while the bars in the background are representing the effective sample size from \cref{eqn:weff}. The correction shifts the free energy downwards and towards larger volumes.}
    \label{fig:SJEOS}
\end{figure}

The absence of imaginary phonon frequencies make it possible to calculate the vibrational contribution to the free energy according to \cref{eqn:vib}. \cref{fig:SJEOS} depicts the volume dependence of the free energy at four different temperatures. 
\cref{eqn:SJEOS}, captures the behavior of c-\ce{HfO_2} across the studied temperature range and as expected the equilibrium volume increases with temperature. Interestingly, $F_\text{corr}$ not only shifts the curve to lower energies, $\mathcal{O}( \SI{100}{\milli\electronvolt\per\formulaunit})$, but it also changes the positions of the minima. As expected, $F_\text{corr}$ gets larger with temperature, i.\,e.\ with increasing anharmonic contributions to the relevant parts of the PES. As a result, the contribution generally favors larger volumes and can be expected to be important for a correct predicting of thermal expansion. 

\begin{figure}[h]
    \centering
    \includegraphics{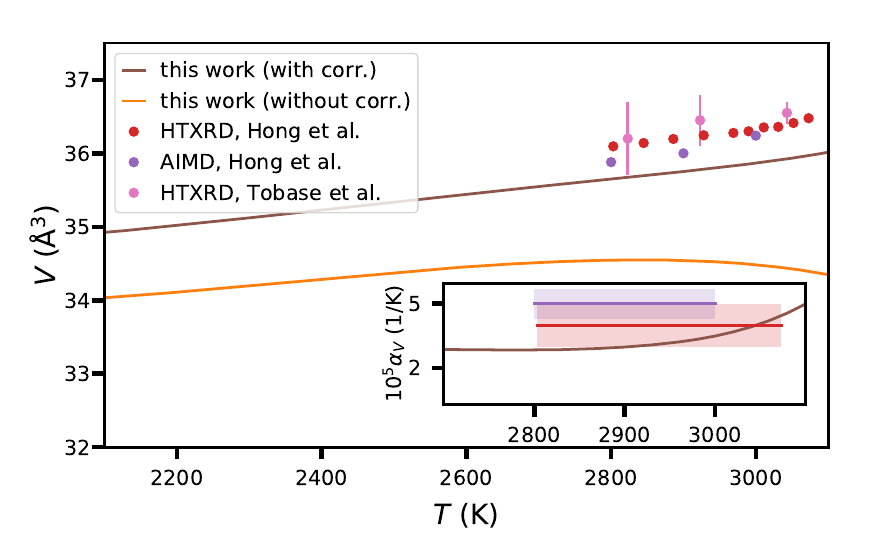}
    \caption{The unit-cell volume expansion as a function of temperature obtained in this work with (without) correction in brown (orange). The results are compared to experimental data \cite{tobase_PRL_2018,hong_nature_2018} and ab-initio molecular dynamics calculations \cite{hong_nature_2018}. The thermal expansion coefficient, $\alpha_V = \frac{1}{V}\left(\frac{\partial V}{\partial T}\right)_p$, is shown in the inset. The shaded area in the inset indicates the experimental uncertainty. }
    \label{fig:thermal_exp_lit}
\end{figure}

The role of the correction term becomes even more apparent when looking at the thermal expansion in \cref{fig:thermal_exp_lit}. 
Comparing the unit-cell volume of c-\ce{HfO_2} with experimental and theoretical results from literature, \cite{hong_nature_2018,tobase_PRL_2018} 
illustrates how neglecting $F_\text{corr}$ will provide underestimated unit cell volume in situations where anharmonicities contribute a significant portion of the total energy. Such situations can arise when describing materials that are inherently anharmonic, or generally for materials at elevated temperatures. \ce{HfO_2} is in our study subject to both of these circumstances and thus any description not taking the anharmonic correction into account is bound to lead to inaccurate conclusions. Even when $F_\text{corr}$ is included, a constant offset of about \SI{0.5}{\angstrom^3}, or \SI{1.2}{\percent}, can be observed. This can be attributed to inherent approximations of the chosen DFT functional. It is not possible to prove this as no \SI{0}{\kelvin} experimental data exists to compare with. However, it is worth noting that the AIMD study reported in Ref.~\onlinecite{hong_nature_2018} (employing the same PBE functional), also finds slightly lower volumes than experiment. 

The calculated thermal expansion coefficient [see inset in \cref{fig:thermal_exp_lit}] is approximately constant, \SI{3.3e-5}{\per\kelvin}, until an increase in thermal expansion can be seen as the melting temperature, $T_m \approx \SI{3100}{\kelvin}$, is approached. The thermal expansion coefficient is not impacted by a small constant offset, and our results are within or very close to the uncertainty, indicated by a red bar, of the experimental average over the range from \SIrange{2800}{3100}{\kelvin}, \SI{4\pm 1 e-5}{\per\kelvin}, obtained by Hong. et al. \cite{hong_nature_2018}.
The volume data from Tobase et.\ al \cite{tobase_PRL_2018} would result in $\alpha_V = \SI{4.39e-5}{\per\kelvin}$, over the region of approximately \SIrange{2823}{3043}{\kelvin}. However, due to the large uncertainty in the volume measurements, the error bar would span about \SI{8.3e-5}{\per\kelvin}, and hence it is not shown in the graph.

Finally, we report the temperature dependence of the c-\ce{HfO_2} bulk modulus.  Due to the fact that the second derivative of \cref{eqn:SJEOS} w.r.t. lattice parameter (or volume) is not linear, the bulk modulus obtained by our method is naturally dependent on the volume at a given pressure at which it is evaluated. Evaluated at \SI{0}{\giga\pascal} we find a bulk modulus of $B_0 = \SI{180}{\giga\pascal}$, which decreases to $\SI{120}{\giga\pascal}$ over the range of \SIrange{2100}{3100}{\kelvin}. This drastic softening is expected as we are approaching the melting point of the material. In their recent study, Irshad et al. \cite{irshad_ACS_2020}, have measured the bulk modulus of pressure-stabilized, nanocrystalline c-\ce{HfO_2} at ambient temperature, finding $B_0 = \SI{242\pm16}{\giga\pascal}$. Applying a small pressure of \SI{4}{\giga\pascal} (corresponding to a volume change of \SI{1.5}{\percent}) to our result, yields a comparable bulk modulus of about \SI{210}{\giga\pascal} at \SI{2100}{\kelvin}, decreasing to \SI{170}{\giga\pascal} at \SI{3100}{\kelvin}. 

\section{Conclusion}{\label{sec:conclusion}}
The behavior of c-\ce{HfO_2} in the high-temperature regime was studied using effective harmonic potentials. At elevated temperatures the unstable mode exhibited by the cubic structure hardens, resulting in a stable phonon spectrum. It was shown that, without consideration of the anharmonic correction term, an accurate description of this phase is not possible and it is conjectured that this term is crucial throughout a broad spectrum of high-temperature materials studies. 

The thermal expansion behavior reported, $\alpha_V=\SI{3.3e-5}{\per\kelvin}$, is in good agreement with the existing experimental and theoretical data, if averaged over the same temperature range. In the range of \SIrange{2100}{3100}{\kelvin} the bulk modulus of c-\ce{HfO_2} exhibits a drastic elastic softening from \SI{180}{\giga\pascal} to \SI{120}{\giga\pascal}, which can be expected as the melting point of the compound is estimated to be around \SI{3100}{\kelvin}.

Ultimately, taking into consideration the difficulties of precise measurements at these high temperatures, as well as the computational cost of the alternatives, effective harmonic potentials can provide valuable insights at manageable cost when studying high-temperature phases.

\section*{Acknowledgements} \label{sec:acknowledgements}
AI4DI receives funding within the Electronic Components and Systems for European Leadership Joint Undertaking (ESCEL JU) in collaboration with the European Union’s Horizon2020 Framework Programme and National Authorities, under grant agreement n° 826060.

\bibliography{biblio.bib}

\end{document}